\documentstyle{mn}

\title[Kerr BH fly-wheel model for statistics of QSOs/AGNs]
{
An Application of Kerr Blackhole Fly-Wheel Model to Statistical Properties of QSOs/AGNs
}
\author[Shin-ya Nitta]
{Shin-ya Nitta $^{1,2}$\thanks{E-mail: snitta@th.nao.ac.jp} \\
$^1$ Division of Theoretical Astrophysics, National Astronomical Observatory Japan, Osawa 2-21-1, Mitaka 181-8588, Japan\\
$^2$ Department of Astronomical Science,The Graduate University for Advanced Studies, Osawa 2-21-1, Mitaka 181-8588, Japan\\ 
E-mail: snitta@th.nao.ac.jp}

\date{ Accepted 1999 May 5.
      Received 1999 April 21;
      in original form 1998 September 4}
\begin{document}
\maketitle
\begin{abstract}

The aim of this work is to demonstrate the properties of the magnetospheric model around Kerr blackholes (BHs), so-called the fly-wheel (rotation driven) model. The fly-wheel engine of the BH-accretion disk system is applied to the statistics of QSOs/AGNs. In the model, 
the central BH is assumed to be formed at $z \sim 10^2$ and obtains nearly maximum but finite rotation energy ($\sim$ extreme Kerr BH) at the formation stage. The inherently obtained rotation energy of the Kerr BH is released through an magnetohydrodynamic process. This model naturally leads finite lifetime of AGN activity. 

Nitta et al. (1991) clarified individual evolution of Kerr BH fly-wheel engine which is parametrized by BH mass, initial Kerr parameter, magnetic field near the horizon and a dimension-less small parameter. We impose a statistical model for the initial mass function (IMF) of ensemble of BHs by the Press-Schechter formalism. By the help of additional assumptions, we can discuss the evolution of the luminosity function and the spatial number density of QSOs/AGNs.

Comparing with observations, somewhat flat IMF and weak dependence of the magnetic field on BH mass are preferred. 
The result well explains decrease of very bright QSOs and decrease of population after $z \sim 2$.

\vspace{0.5cm}


\end{abstract}

\vspace{1cm}

\section{INTRODUCTION}

We discuss the evolution of QSO/AGN activities under the fly-wheel (rotation driven) model which is one of the plausible models for the powerful engine of the AGNs including a rotating central blackhole (BH). This fly-wheel engine might not be familiar comparing with the fuel engine (accretion driven engine), however, this is very attractive because this model can explain the evolution and the lifetime of AGN activities very naturally. 

It is widely believed that recent discovery of the red tail of emission lines (Fe K$\alpha$) from the central region of AGNs suggests that the central blackholes are quickly rotating, i.e., the monster BH should be the Kerr BH (see Tanaka et al. 1995, Iwasawa et al. 1996, Dabrowski et al. 1997). The innermost stable circular orbit around the Kerr BH can intrude more close to the horizon than the Schwartzschild hole (non-rotating BH) of the same mass. Hence the line emission from the innermost region of the accretion disk should be considerably red-shifted by the gravitation, and makes the red tail. 

From the theoretical point of view, the majority believes that the central BH presumably gets an enormous angular momentum at the formation stage of the monster. For example, Sasaki \& Umemura (1996) showed the formation scenario by using the Compton drag process as follows. After the neutralization of universe at $z \sim 10^3$, density fluctuations grow to form the proto-galactic cloud, and each fluctuation obtains the angular momentum through the tidal interaction among them. At the era $z \sim 10^2$, nuclear reactions occur in the central region of the rotating proto-galactic cloud, and the matters are reionized by stellar UV radiation. The rotating reionized matters must interact again with the uniformly distributed cosmic background radiation from the last scattered surface. This interaction efficiently extracts the angular momentum of the cloud material to the radiation field by the Compton scattering, then the angular momentum of the matter decreases, and when it goes just below a critical value of the angular momentum, the matter suddenly collapses and makes the quickly rotating monster blackhole. 

In this scenario, the initial angular momentum of the central hole shall be very close to the maximum value (say, $a \sim m$ where $a$ is the Kerr parameter and $m$ is the mass of the hole) which the Kerr hole can hold. Hence it seems reasonable to suppose that the central blackhole is similar to the extreme Kerr hole at the formation stage. Such holes have an enormous rotation energy ($\sim 10^{54} \mbox{J} \sim 10^{39} \mbox{W} \times 10^9 \mbox{yr}$ for the hole with mass $\sim 10^8 M_\odot$). This is enough to explain the total energy release of AGNs. 

There are two different types of engines for energy production at the blackhole-accretion disk systems. The first is the well-known ``fuel engine'', which is the accretion powered engine. This is the major one which has been frequently adopted to explain the AGN activities. The fuel engine acts by a process to convert the gravitational energy released from the infalling matter to the radiation. The standard disk model (Shakura \& Sunyaev 1973) is the representative of this model. 

The second is the ``fly-wheel engine''. This is the rotation powered engine. While the fly-wheel engine is not so familiar in the field of AGNs, this engine is as powerful as the fuel engine, and has very interesting features as discussed in the following sections. In contrast with the fuel engines, the energy source is the rotation of the Kerr BH itself. Of course, the rotation of the accretion disk also can be another energy source. However our scope is focused to the case in which the rotation of the BH is energy source, because, the rotation energy of the disk is supplied by accretion, so that there is an apprehension of confusion of two engines. The author strongly hopes to introduce this fascinating engine to researchers working in the field of AGNs. The comparison of the fly-wheel engine and the fuel engine is discussed in section 2.

Let us summarize the properties of the fly-wheel engine. The idea to extract the rotation energy of the Kerr holes is firstly proposed by Penrose (1969). When an incident particle into the Kerr hole splits into two parts inside the ergo-sphere with a very large relative velocity, one particle can be thrown into the negative energy orbit falling to the hole and another particle escapes outward with larger energy than the initial energy of the incident particle. In this case, the rotation energy of the hole is reduced by the infall of the negative energy particle, and reduced energy is carried by the escaping particle. This is well-known ``Penrose process''. Unfortunately it is pointed out that the Penrose process is not effective for astrophysical problems (see Bardeen et al. 1972) because the critical value of the relative velocity in order to realize the negative energy orbit is close to one half of the light velocity. Such a large relative velocity can be realized only by nuclear reactions of particles and may not be achieved by usual dynamical processes, e.g., the tidal disruption of the accreting matter. 

The electromagnetic mechanism extracting the rotation energy of the Kerr BHs is firstly proposed by Blandford \& Znajek (1977). This is well known as the ``magnetic breaking process'' or the ``BZ process''. In their study, the magnetosphere is supposed to be ``force-free'' (strictly saying, this is ``magnetically dominated''), and they clearly showed the energy extraction in the form of the Poynting flux when the rotation speed of BH is greater than it of the magnetosphere. An extension of the magnetic breaking process to the full MHD (magnetohydrodynamic) system was performed by Takahashi et al. (1990) as an elementary process of the BH engines. By precise analysis of the MHD accretion flow onto the Kerr BH, they succeeded to clarify the condition to realize the ``negative energy MHD inflow''. This process is named as the ``MHD Penrose process''. Nitta et al. (1991) studied the magnetospheric structure filled with the trans-magnetosonic MHD inflow onto the Kerr hole, and applied the MHD Penrose process to the problem of the individual evolution of AGNs. The result of this work is briefly reviewed in section 3. 

Recently, the BZ process is speculatively staged again as the elementary process of the $\gamma$-ray burst (GRB, see Paczynski 1998). In this case, the rotation energy $\sim 10^{47}$ [J] of nearly maximum rotating Kerr BH of mass $\sim 10 M_\odot$ is considered to be extracted by very strong magnetic field $\sim 10^{11}$ [T] in a few seconds. The extracted Poynting energy is expected to produce the ultra relativistic wind with the Lorentz factor $\mathrel{\hbox{\rlap{\hbox{\lower4pt\hbox{$\sim$}}}\hbox{$>$}}} 10^2$. Of course, ``magnetically dominated'' assumption of original BZ process is too simple to treat the wind acceleration, and it should be extended to full MHD fly-wheel model. The fly-wheel model is still unclear, but it should be one of fascinating process to unify physics of quasars and micro-quasars. 

\begin{figure}
\unitlength=1mm
\vspace{8cm}
\includegraphics{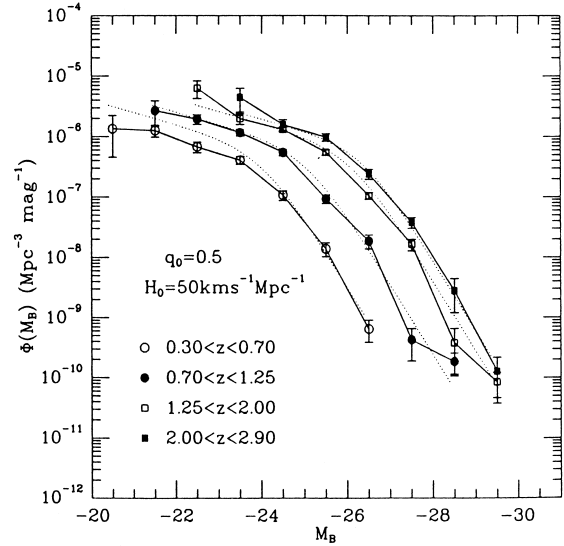}

\caption{Observed luminosity function of QSOs from Boyle et al. (1991). QSO LF for $z<2.9$ in a $q_0=0.5$ universe. The error bars are based on Poisson statistics. The dotted lines indicate the derived model fit to the data. \label{fig:LFO}}
\end{figure}

\begin{figure}
\unitlength=1mm
\vspace{8cm}
\includegraphics{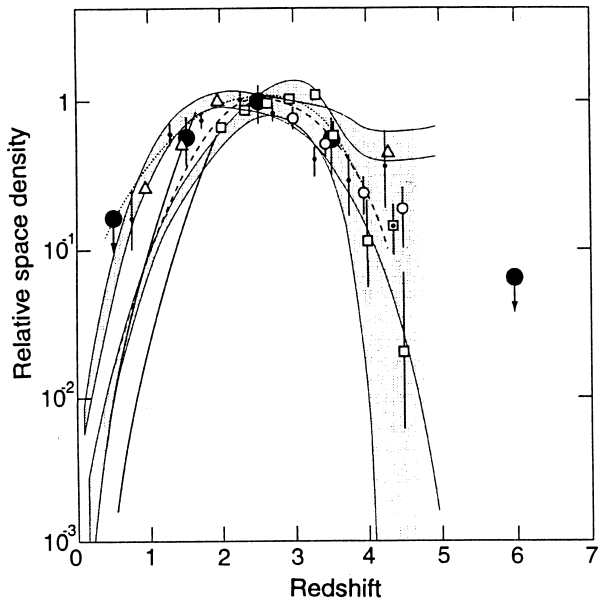}
\caption{Observed evolution of QSO population from Shaver et al. (1996). Space densities, comoving, normalized to $z=2 \sim 3$ are plotted as a function of red shift for flat-spectrum radio-loud quasars with $\geq 1.1 \times 10^{27}$[W Hz$^{-1}$ sr$^{-1}$] (large filled circles). The optically selected quasar samples of Warren et al. (empty squares), Schmidt et al. (empty circles), Kennefick et al. (dotted square), Hawkins \& V\'{e}ron (small filled circles), and Irwin et al. (combined with Miller et al. (empty triangles)), the flat and steep spectrum sources studied by Dunlop \& Peacock (shaded areas), and the flat spectrum radio-loud quasar sample of Hook et al. (dotted line). \label{fig:popO}}
\end{figure}

In this paper, the result of Nitta et al. (1991) for individual evolution of fly-wheel engine is applied to the statistics of ensemble of QSOs/AGNs and compared with the observation. Figure \ref{fig:LFO} shows observation of the luminosity function (LF) of QSOs. Figure \ref{fig:popO} shows observation of the evolution of the spatial number density of QSOs. Our attention will be focused to explain the evolution of QSOs/AGNs in a range $0 \leq z \leq 5$ by a mechanical process. 

In order to discuss the physical process of the plasma inflows and the magnetospheric structure of the Kerr BH, we suppose general relativistic, stationary and axisymmetric ideal cold MHD flows. In this case, MHD equations reduce to well-known basic equations: the Bernoulli equation and the Grad-Shafranov equation (see Takahashi et al. 1990 and Nitta et al. 1991) with constants of the motion. By using these basic equations, we discuss the properties of the fly-wheel engine and apply it to the evolution of ensemble of QSOs/AGNs in section 4. We should note again that our primary purpose is to demonstrate the fascinating properties of the fly-wheel model and not to produce a serious model for evolution and statistics of QSOs/AGNs.

\section{COMPARISON OF FLY-WHEEL ENGINE VS. FUEL ENGINE}

Here we compare the fly-wheel model with the fuel model, and clarify the differences between them. Our attention is focused to the energy source, the power output and the form of the energy transfer. 

The most fundamental difference is in the energy source of them. The energy source of the fuel engine is gravitational energy of infalling matter released through some dissipation process like $\alpha$-viscosity (Shakra \& Sunyaev 1973). Hence this engine can act while the accretion is continued. The energy source of the fly-wheel engine is the rotation energy of the central spinning BH itself which can be extracted through an electromagnetic process (the magnetic breaking). We should note that the rotation energy of the BH is obtained at the formation stage, and is finite. Thus the lifetime of the fly-wheel engine, on the contrary, must be finite. 

The output power of the fuel engine essentially depends upon the mass accretion rate of the infalling matter, and is widely variable. Upper boundary of the power approximately corresponds to the Eddington luminosity. On the contrary, the output power of the fly-wheel engine is determined by the magnetospheric equilibrium. In a typical case, the power $\sim 10^{39}$[W] for the BH mass $\sim 10^8$ M$_\odot$ (see the next section), and is enough to explain actual QSOs/AGNs. 

In the fuel engine, generated power is in a thermal form (e.g., the standard disk model) and is immediately converted to the radiation from the central region. The mechanical process of the fueling is so complicated. We need several models of the angular momentum extraction for each decade of distance from the BH. These mechanisms must be matched consistently, however, this is very difficult. 

In the fly-wheel engine, the extracted rotation energy from the BH is once stored in the form of the Maxwell stress of the magnetosphere. This stress causes the global electric current circuit in the magnetosphere, and magnetocentrifugal force drives the plasma outflow, for example, the highly collimated bipolar jets in radio-loud AGNs or the equatorial wind in BAL QSOs. The kinetic energy of the plasma outflow is finally converted to radiation at some distant region through some emission processes (e.g., the synchrotron radiation produced by the 1st Fermi acceleration on the shock). The mechanical process of the fly-wheel engine is simple. We can clarify it by closed discussion in the vicinity of the BH. In addition, we should note that the fly-wheel model can be a simple and unified mechanism throughout the entire magnetosphere in a range from AU to Kpc or Mpc. 

These properties of the fuel engine and the fly-wheel engine are summarized in table 1. 

\begin{table*}
\begin{minipage}{160mm}
\caption{Comparison of Fly-Wheel Model vs. Fuel Model \label{tab:FF}}
\label{mathmode}
\begin{tabular}{|c|c|c|}\hline
Name& Fuel&Fly-Wheel \\ \hline
Energy Source& Gravitational energy of accreting matter&Rotation energy of Kerr BH \\ \hline
Lifetime&Eternal &Finite Lifetime \\ \hline
Energy Transfer& Thermarized at central region  & Induces the magnetospheric stress\\
 &$\Rightarrow$Radiation&$\Rightarrow$Plasma outflow propagating toward distant region\\ \hline
Applicable Region&Central activity &Central($\sim$AU) to Distant($\sim$Mpc) Region \\ \hline
Properties& Needs different fueling mechanisms  & Closed discussion around BH is possible\\ 
&for each decade of distance& Simple \& unified scheme\\ \hline
\end{tabular}
\end{minipage}
\end{table*}

\section{EVOLUTION OF THE POWER OUTPUT FROM THE KERR BLACKHOLE MAGNETOSPHERE}

In the MHD scheme, the magnetosphere near the horizon of a BH must be filled with super-magnetosonic accretion flow to keep the causality. The trans-magnetosonic condition crucially restricts the magnetospheric structure. 

Nitta et al. (1991) studied the magnetospheric structure of a Kerr BH filled with trans-magnetosonic accretion flow. The BH magnetosphere is characterized by coexistence of the outgoing flow and the accreting flow. 
In order to realize the outgoing flow and the accreting flow@simultaneously, they suppose the stagnation region (source region) which is sustained by the magnetocentrifugal force against the gravity. This is the source of ingoing/outgoing flows. The stagnation region may correspond to the pair creation region near the outer gap (near the surface $\omega=\Omega_F$ where $\Omega_F$ is the angular velocity of the magnetosphere and $\omega$ is the Lense-Thirring rotation of the inertial frame: see Hirotani \& Okamoto 1998) or the disk halo. The accretion flow starts with very low poloidal velocity, and accelerates toward the horizon, then the flow must pass through the Alfv\'{e}n point and the fast point before reaching the horizon. 

In their result, it is clarified that the strong gravity of BH cause the accretion flow and amplifies the magnetic flux, but the total magnetic flux $\Psi_H$ and the particle number flux $\dot{N}_H$ threading the horizon are suppressed by the rotation of the hole,  
\begin{equation}
\Psi_H \sim \frac{\Omega_F}{\omega_H} \Psi_0
\end{equation}
where $\Psi_0$ denotes the total magnetic flux of the magnetosphere and $\omega_H$ is the angular velocity of the Kerr blackhole (the Lense-Thirring angular velocity at the horizon), 
\begin{equation}
\dot{N}_H \sim \frac{{B_0}^2}{\mu \Omega_F \omega_H}
\end{equation}
where $B_0$ is the magnetic field at the source region of the accretion flow and $\mu$ is the averaged rest mass of the particle of the flows. We also find that one infalling particle can release the rotation energy of BH of the order of its rest mass energy. 

According to these results we can estimate the total power output $L_{BH}$ from the rotating BH as
\begin{equation}
L_{BH}(m, \epsilon, B_0; t)=P_0(m,B_0,\epsilon) \cdot \frac{\theta(t-t_{max}(m,B_0,\epsilon))}{\sqrt{1-t/\tau_{evo}(m,B_0,\epsilon)}} \label{eq:LBH}
\end{equation}
where $t$ is the time after the birth of BH, $P_0=m^2 B_0^2/\epsilon$, 
$\tau_{evo}=\epsilon^2/(m B_0^2)$, 
$t_{max}=(1-\epsilon^2) \tau_{evo}$, 
$\epsilon \equiv m \Omega_F$ (always less than unity, see Nitta et al. 1991), 
$\theta$ is the Heaviside function ($\theta(x)=0$ for $x<0$, $\theta(x)=1$ for $x \geq 0$). We should note that this formula is somewhat simplified from the original form (see equation [5.9] of Nitta et al. 1991). Since this formula is derived by estimation of order of magnitude, the formula contains unspecified factors of order of unity. Here we assume these factors to be unity for simplicity. For the typical case, the values of $P_0$ and $\tau_{evo}$ are given as 
\begin{equation}
P_0 \sim 10^{39} [\mbox{W}] \left(\frac{\epsilon}{0.1}\right)^{-1} \left(\frac{m}{10^8 M_\odot}\right)^2 \left(\frac{B_0}{1[\mbox{T}]}\right)^2\ ,
\end{equation}
\begin{equation}
\tau_{evo} \sim 10^9[\mbox{yr}] \left(\frac{\epsilon}{0.1}\right)^2 \left(\frac{m}{10^8 M_\odot}\right)^{-1}\left(\frac{B_0}{1[\mbox{T}]}\right)^{-2}\ .
\end{equation}
In this model, initially quickly rotating BH ($\omega_H \gg \Omega_F$) spins down by the magnetic breaking process and releases its rotation energy. When $t=t_{max}$ ($\omega_H=\Omega_F$), output power is maximum, then the engine ceases to act suddenly. 

\begin{figure}
\unitlength=1mm
\vspace{8cm}
\includegraphics{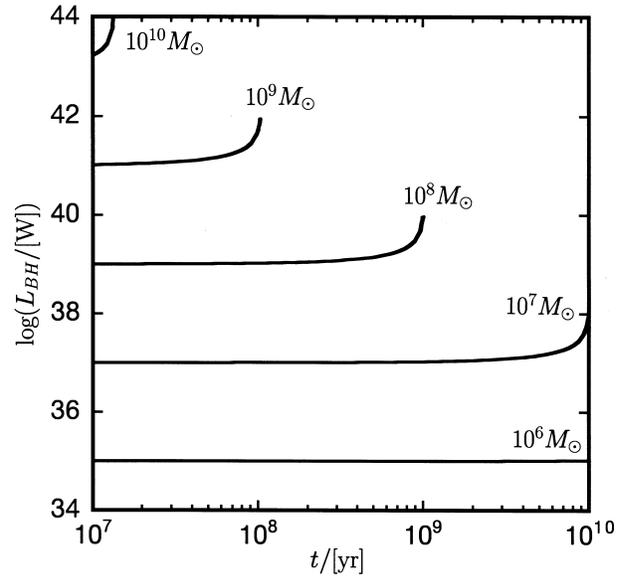}
\caption{Light curves of the fly-wheel engine for different BH mass, $m=10^6 M_\odot$, $10^7 M_\odot$, $10^8 M_\odot$, $10^9 M_\odot$, $10^{10} M_\odot$ for $B_0=1$[T] and $\epsilon=0.1$. BHs with $m \sim 10^8 M_\odot$ cease to act at $z \sim 3$ and important for evolutionary properties at this range. \label{fig:LBH}}
\end{figure}

A typical sample is shown in figure \ref{fig:LBH} where $L_{BH}$ is denoted as a function of the time after the BH formation. In this scheme, the extracted energy is in the form of the Poynting flux, and it will be converted to the thermal energy of the magnetospheric plasma through some dissipative process or the kinetic energy of outflows by magnetocentrifugal drive which will act outside the source region. Let us assume here that all the extracted energy converted to the radiation through some unspecified mechanism, hence the total output power should be interpreted as the bolometric luminosity. We should note that during the evolution, the BH mass $m$ is nearly constant, because the time scale of the mass variation is much longer than it of the angular momentum variation which determines the time scale of evolution (see Nitta et al. 1991). 

Here, the evolution of power output of individual Kerr BH magnetosphere has been clarified. We now try to apply this result to the statistical discussion for ensemble of QSOs/AGNs. 

It should be noted that the strength of the BZ process depends crucially on that of the magnetic field threading the horizon. Blandford \& Znajek (1977) firstly discussed the BH fly-wheel engine, but total magnetic flux on the horizon is a free parameter in their discussion based on the magnetically dominated limit. On the contrary, Nitta et al. (1991) using full MHD discussion enable to obtain the magnetic flux on the horizon as a result of the inner magnetospheric equilibrium based on full MHD. Instead of it, the total magnetic flux of the entire magnetosphere is treated as a free parameter. This point will be discussed in section 6.

\section{APPLICATION TO STATISTICS OF QSOs/AGNs}

General relativistic theory of QSO core formation is still an open question. Hence we do not have statistical properties of the parameters of QSO BHs. If we assume the statistical distributions of the BH mass $m$ (the initial mass function), the Kerr parameter $a$ (the initial Kerr parameter function) of seed BH and the magnetic field strength $B_0$ at the source region (this should depend on the BH mass, the accretion rate and the dynamo theory), we can sum up the contribution of each BH over the ensemble, and can suggest the statistical properties of QSOs/AGNs by the Kerr BH fly-wheel model. 

Here we will demonstrate a preliminary application of Kerr BH fly-wheel model to QSO statistics. The discussion is based on the Press-Schechter formalism as a probable seed BH formation scenario. Sasaki \& Umemura (1996) discussed an additional process, the Compton drag scenario, for further angular momentum extraction to form the proto-galactic cloud to form the seed BH, and suggest the initial mass function in figure 1 of their paper. Unfortunately, the distribution of the Kerr parameter and the magnetic field strength at the source region are not mentioned there. Hence we should assume the magnetic field strength $B_0$ at the source region and the initial Kerr parameter $a/m$ as follows. 

 The magnetic filed at the source region is usually estimated as $B_0 \sim 1$[T] for $m=10^8 M_\odot$ in order to explain typical QSO luminosity. This value probably depends on the BH mass, then we assume 
\begin{equation}
B_0 = 1 [\mbox{T}] \times (m/10^8 M_\odot)^\zeta \ . \label{eq:B0}
\end{equation}
We also assume the initial Kerr parameter $a/m \sim 1$ (nearly the extreme Kerr BH at the initial stage) and the small parameter $\epsilon \equiv m \Omega_F \sim 0.1$. Then the power output $L_{BH}$ is the function of $m$ and $t$ (see  equation \ref{eq:LBH}). Thus the only we need is the initial mass function of BHs.

\subsection{Evolution of the luminosity function}

From Sasaki \& Umemura (1996) we obtain the initial mass function $f_{BH}$ of BHs based on the standard CDM model as 
\begin{equation}
f_{BH}(m)=\frac{n+3}{6}\frac{\rho_0}{M(m)^2}\sqrt{\frac{2}{\pi}}\nu e^{-\nu^2/2} \label{eq:IMF}
\end{equation}
where 
\begin{equation}
\nu=\left(\frac{M(m)}{M_{c0}}\right)^{(n+3)/6} (1+z_{vir})\ ,
\end{equation}
with $M$ is the total (dark matter+baryon) mass of the proto-galactic cloud, $\rho_0$ is the present total mass density, $M_{c0}=\rho_0 4 \pi (16 \mbox{Mpc})^3/3$. $M$ is related with the BH mass $m$ as 
\begin{equation}
m=r_{BH} \Omega_b M\ ,
\end{equation} 
where $\Omega_b$ is the fraction of the baryonic mass to the total mass and $r_{BH}$ is the ratio of the BH mass to the bakoryonic mass. We assume $\rho_0=6.9 \times 10^{10} [M_\odot/\mbox{Mpc}^3]$, $\Omega_b=0.05$ and $r_{BH}=0.1$ in this paper (we adopt a cosmological model with total density parameter $\Omega_0=1$ and the present Hubble constant $H_0=50$[km/s/Mpc]). 

\begin{figure}
\unitlength=1mm
\vspace{9cm}
\includegraphics{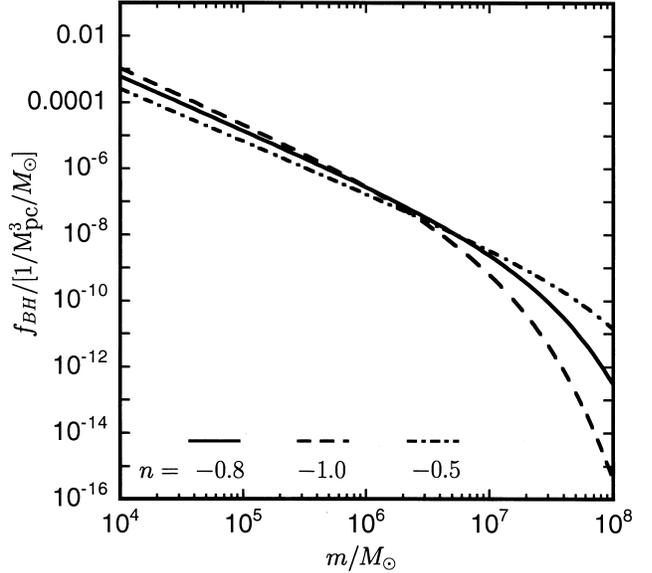}
\caption{Initial mass functions of BHs for $n=-1.0$ (dashed line), $-0.8$ (solid line), $-0.5$ (dot-dash line). As a result, the case $n=-0.8$ gives preferable evolution of the luminosity function and the spatial number density of QSOs. \label{fig:IMF}}
\end{figure}

In their paper, the BH formation epoch $z_{vir}$ is obtained from somewhat complicated procedure, however, in a BH mass range $10^6 M_\odot \leq m \leq10^{10} M_\odot$, $z_{vir}$ varies in a very narrow range around $z_{vir} \sim 200$, then we neglect the mass dependence for simplicity and put $z_{vir} \equiv 200$ throughout this paper. The resultant mass functions of BHs are shown in figure \ref{fig:IMF}. 

We can obtain the luminosity function $\Phi$ at the cosmic time $t$, 
\begin{eqnarray}
\Phi(m; t)&=\left|\frac{dn_{BH}(m)}{dL_{BH}(m; t)}\right| \nonumber\\
               &=\left|\frac{dn_{BH}(m)}{dm}/\frac{dL_{BH}(m; t)}{dm}\right| \nonumber\\
               &=\left|\frac{f_{BH}(m)}{dL_{BH}(m; t)/dm}\right|  \label{eq:Phi},
\end{eqnarray}
where $n_{BH}(m)$ is the total number of BHs having the mass smaller than $m$. Usually, evolution of the luminosity function is parametrized by $z$ instead of $t$. We adopt here the Einstein-de Sitter universe as the cosmological model to relate $t$ with $z$, 
\begin{equation}
t=t_0/(z+1)^{3/2} \label{eq:t-z}
\end{equation}
where $t_0 \sim 10^{10}$[yr] is the present time. 

Here we discuss a number of examples of the dependence of the magnetic filed $B_0$ at the source region on BH mass $m$. The locus of the plasma source is supposed to several times the horizon radius ($\sim m \epsilon^{-2/3} \sim 4.6 m$ for $\epsilon=0.1$) where the pair creation seems to be effective due to the outer gap model (Hirotani \& Okamoto 1998). The case $B_0 \propto m^{-1/2}\ (\zeta=-1/2)$ is so-called the Eddington value. This formula is derived as follows. Based on the spherical accretion with the Eddington accretion rate, we suppose the equipartition condition of energy density between the gravitational one and the magnetic one. However, in this case, the lifetime of the fly-wheel engine is determined independent of $m$ from equation (\ref{eq:LBH}). We can easily imagine the evolution of the luminosity function of this case. The curve moves rightward holding its shape corresponding to the increase of output power, and arrives at the explosive stage, and then, all the BH engines simultaneously cease their activity. This is trivially inconsistent with the observation (see figure \ref{fig:LFO}), hence we reject this case. 

The magnetic field of the inner region of the Shakura-Sunyaev's accretion disk is evaluated as follows. They assumed the equipartition between the magnetic energy and the thermal energy. For the radiation pressure supported case, $\zeta = -1/2$ (similar to the Eddington value). This case is also unsuitable as discussed above. For the gas pressure supported case, $\zeta = -1/20$ assuming that the mass accretion rate is proportional to $m$ (like the Eddington limit). In our discussion, since the dynamic range of the BH mass $m$ is $10^6 M_\odot \leq m \leq10^{10} M_\odot$, merely in the dynamic range of the fourth order of magnitude, the dependence $\zeta = -1/20$ means almost $B_0 \sim const.$ Hence we can assume that the mass dependence of the magnetic field $B_0$ at the source region is presumably very weak, $-1/2 < \zeta$. In the following discussion, we assume $\zeta=0$ and set $B_0=1$[T] independent of $m$. 

\begin{figure}
\unitlength=1mm
\vspace{8cm}
\includegraphics{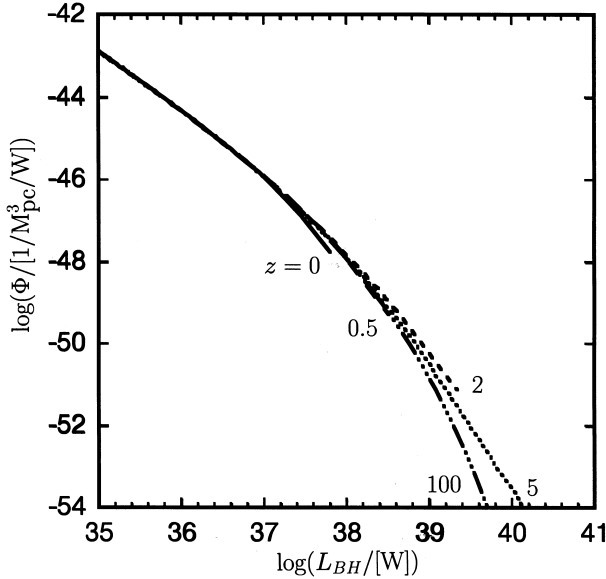}
\caption{Evolution of luminosity function of the fly-wheel engine for $\zeta=0$ and $n=-0.8$. The brighter-end of the curve once lifts up for $5>z>2$ then it drops and shrinks for $2>z>0$. This tendency is consistent with observations. \label{fig:LF1}}
\end{figure}

\begin{figure}
\unitlength=1mm
\vspace{8cm}
\includegraphics{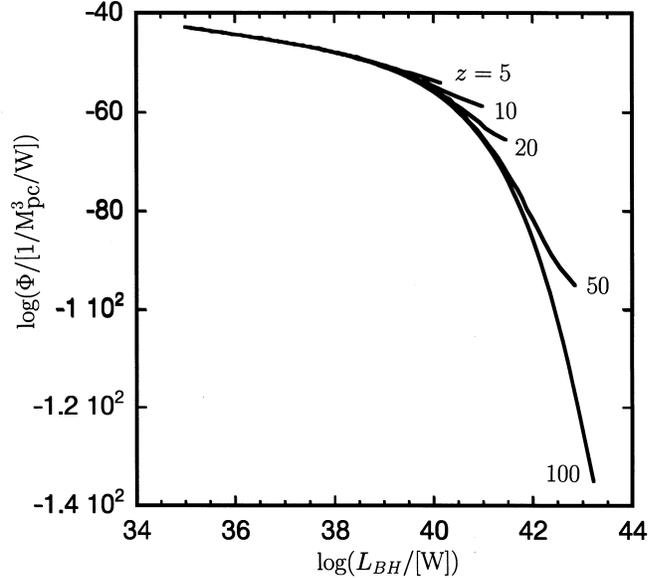}
\caption{Evolution of luminosity function of the fly-wheel engine for $\zeta=0$ and $n=-0.8$ (zoom-out) in the period $100>z>5$. The brighter-end of the curve considerably lifts up and shrinks by evolution. \label{fig:LF2}}
\end{figure}

Evolution of the luminosity function is shown in figure \ref{fig:LF1}, \ref{fig:LF2} for the typical case $\zeta=0, n=-0.8$. 
In case $\zeta=0$, the initial luminosity is proportional to $m^2$, hence the luminosity function at the formation stage $z \sim z_{vir}$ remarkably reflects the initial mass function (see figure \ref{fig:IMF}). The curve of the luminosity function consists of monotonically decreasing power-law slope and exponential cut-off. 

For $\zeta=0$, the lifetime of the fly-wheel engine is a decreasing function of $m$ (see equation \ref{eq:LBH}). When the BHs of mass, say, $m=m_0$ approach to the explosive stage $t=t_{max}$, the luminosity $L_{BH}(m_0,t)$ and $dL_{BH}(m; t)/dm|_{m=m_0}$ quickly increase. Until this time, the BHs having mass $m > m_0$ have already died. Hence the brighter-end of the luminosity function quickly extended in the lower-right direction (see eq. [\ref{eq:Phi}]). If the initial luminosity function has a sufficiently steep sloop, the evolution results a lift up at the brighter-end. On the other hand, if the initial luminosity function has a flat sloop, at the first, the brighter-end lifts up, however, the brighter-end reach the junction point between the power law part and the exponential cut off part of the luminosity function, the initial curve is bend in lower direction by the evolution (see figure \ref{fig:LF1} for $n=-0.8$). This behavior seems to be plausible to explain the observed evolution of the luminosity function shown in figure \ref{fig:LFO} qualitatively. The criterion of the steepness is $n \sim 0.8$ for $\zeta=0$. 

This evolution is just the effect of individual evolution of each fly-wheel engine. The time scale of the evolution is a decreasing function of the BH mass $m$, and the individual luminosity is an increasing function of $m$. Hence, the brighter-end corresponds to massive and short lifetime BHs. Massive BHs quickly evolve to the explosive stage (lift up the curve at the exponential cut off part or more steepen at the power law part), and then they cease to release energy (shorten the curve). 

From the observational studies of the QSOs number counting, the bending of luminosity function curve has been pointed out (see, e.g., Boyle 1993, Pei 1995). The functional form of the curve is guessed as double power law or power law with exponential cut-off. Boyle (1993) argued the evolutionary motion of bending point (see figure 2 of Boyle 1993). The result of the fly-wheel model suggests that the bending point is determined by the initial mass function and does not move during evolution.

\subsection{Evolution of the QSO population}

Similarly we can discuss the evolution of QSO spatial number density (or usually called as ``population'') as a function of $z$. At the time $t$, the BHs of mass $m$ with $t \leq t_{vir}+t_{max}(m)$ are still alive. This condition gives the upper boundary $m_u(t)$ of the mass of the active BHs, because $t_{max}(m)$ is an increasing function of $m$ in case $B_0 \propto m^0$. From the condition 
\begin{equation}
t_{vir}+(1-\epsilon^2) \tau_{evo}(m) \geq t
\end{equation}
we obtain 
\begin{eqnarray}
m_u(t)=&\frac{\epsilon^2 (1-\epsilon^2)}{B_0^2}\frac{1}{t-t_{vir}}\\
           =&10^8[\mbox{M}_\odot] \left(\frac{\epsilon}{0.1}\right)^2 \left(\frac{B_0}{1\mbox{T}}\right)^{-2} (1-\epsilon^2) \left(\frac{10^9 \mbox{yr}}{t-t_{vir}}\right)
\end{eqnarray}
Here after we treat only the active BHs ($m \leq m_u(t)$). 

Next we assume the detection limit $L_{lim}(t)$ of observation on the bolometric luminosity, and let us count the number of BHs satisfying 
\begin{equation}
L_{BH} \geq L_{lim} \label{eq:Llim}
\end{equation}
as QSOs. Note that instead of that we cannot discuss the spectrum of released energy, we can treat only the total output power of the fly-wheel engine. The released energy by the fly-wheel engine produces, at the first, the plasma outflows, and finally, it is supposed to be converted to the radiation through some physical processes like the Fermi acceleration on the shock surface. Hence our attention should be focused to the bolometric luminosity under an assumption that released energy is completely converted to the radiation. Of course, actual observational segregation of QSOs from other objects is based on multicolor spectroscopy, but, unfortunately we cannot argue any more than the bolometric luminosity here. 

Models of concrete functional form of $L_{lim}(t)$ will be given later. The condition (\ref{eq:Llim}) gives the lower boundary $m_l(t)$ of the BH mass which can be detected as QSOs at the time $t$. The relation $L_{BH}(m; t)=L_{lim}(t)$ reduces to an equation for $m$, 
\begin{equation}
B_0(m)^4 m^4/\epsilon^2+L_{lim}(t)^2 t B_0(m)^2 m/\epsilon^2-L_{lim}(t)^2=0 \ .
\end{equation}
This equation can be easily solved numerically. Especially, for case $\zeta=0$ (see equation \ref{eq:B0}), this equation can be reduced to the 4th order algebraic equation, 
\begin{equation}
k_1 m^4+k_2(t) m+k_3(t)=0
\end{equation}
where 
\begin{eqnarray}
k_1&=&B_0^4/\epsilon^2 \\
      &=&10^{60}[\mbox{erg}^2/\mbox{s}^2/\mbox{M}_\odot^4] \left(\frac{\epsilon}{0.1}\right)^{-2} \left(\frac{B_0}{1\mbox{T}}\right)^4 ,
\end{eqnarray}
\begin{eqnarray}
k_2(t)=&L_{lim}(t)^2 t B_0^2/\epsilon^2 \\
=L_{lim}(t)^2& \left(\frac{t}{1 \mbox{yr}}\right) 10^{-17} [1/\mbox{yr}/\mbox{M}_\odot]\left(\frac{\epsilon}{0.1}\right)^{-2} \left(\frac{B_0}{1\mbox{T}}\right)^2 , 
\end{eqnarray}
and $k_3(t)=-L_{lim}(t)^2$. We can obtain unique real-positive root of this equation as $m_l(t)$. Let us sum up the number of BHs in the range $m_l(t) \leq m \leq m_u(t)$, 
\begin{equation}
n_{QSO}(t)=\int_{m_l(t)}^{m_u(t)} f_{BH}(m) dm .
\end{equation}
This is the spatial number density of QSOs. 

\begin{figure}
\unitlength=1mm
\vspace{8cm}
\includegraphics{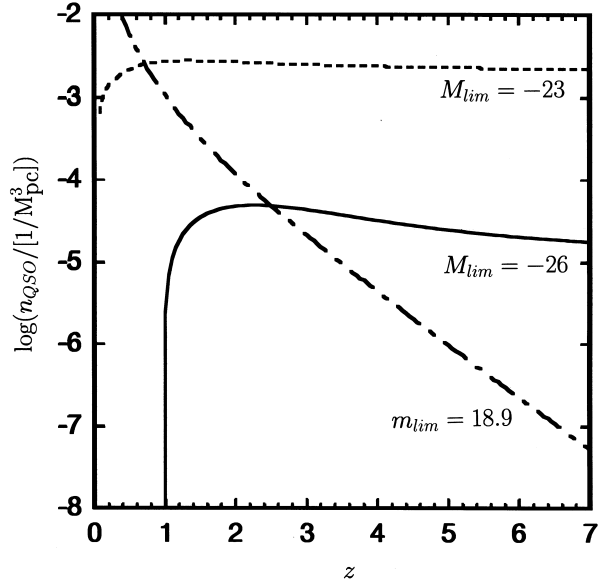}
\caption{Evolution of QSO population for $n=-0.8$. QSOs with bolometric luminosity greater than $L_{lim}$ ($=10^{38.9}$[W] for solid line, $10^{37.7}$ [W]  for dashed line) are counted. Decrease of the population after the peak $z \sim 2$ for solid line is clearly shown as a result of the evolutionary scenario of Kerr BH fly-wheel engine. The dot-dash line denotes the population of QSOs with the energy flux $F \geq 6.6 \times 10^{-16}$[W/m$^2$]. Decrease of the population at high $z$ is more significant than above cases. \label{fig:pop1}}
\end{figure}

The results of simple cases $L_{lim}(t)$ as 
\begin{equation}
L_{lim}(t)=10^{38.9} \mbox{[W] (solid line)} \label{eq:Llimc}
\end{equation}
 and 
\begin{equation}
L_{lim}(t)=10^{37.7} \mbox{[W] (dashed line)}
\end{equation}
 are shown in figure \ref{fig:pop1}. These values correspond to the luminosity at absolute magnitude $M=-26$ (typical value for QSOs) and $-23$ (typical value for AGNs), respectively. 

Figure \ref{fig:pop1} for $n=-0.8$ shows very plausible evolution consistent with observations, in a qualitative sense, but the plot for $n=-1$ (unplotted as figure) shows rather sudden decrease after the peak ($z \leq 3$). This is due to the difference between these cases of evolution of the luminosity function in this range of $z$. 

In the actual observation, the detection limit $L_{lim}$ should be an increasing function of the look back time $t_0-t$ or the red shift $z$, because the detection limit corresponds to the limit on the energy flux $F$ like $F \geq F_{lim}$. We have also tried more realistic function of the detection limit,  
\begin{equation}
L_{lim}(z)=F_{lim} 4 \pi [\sqrt{1+z}-1]^2 (1+z)/(H_0/c)^2 \label{eq:Llimc2}
\end{equation}
where $F_{lim}=$ say $6.6 \times 10^{-16}$[W/m$^2$] (apparent magnitude $m_{lim}=18.9$) is the detection limit on the energy flux, $H_0=50$[km/s/Mpc] is the Hubble constant at the present epoch and $c$ is the speed of light. This is the formula of the translation of the energy flux $F_{lim}$ to the luminosity $L_{lim}$ based on the Einstein-de Sitter universe. The result is plotted as the dot-dash line in figure \ref{fig:pop1}. This value of $F_{lim}$ is a tentative and artifical one chosen to intersect the solid line and the dot-dash line in $z=2.5$.

The conversion from the detection limit on the flux $L_{lim}$ to the limit on the bolometric magnitude $m_{lim}$ is based on the relation that 
\begin{equation}
m_{lim}=-2.5 \log \frac{F_{lim}}{F_0} , 
\end{equation}
where $F_0=2.48 \times 10^{-8} [\mbox{W/m}^2]$ (see Allen 1973). The value of the bolometric magnitude $m_{lim}=18.9$ adopted here might be somewhat brighter than the limit in the magnitude of actual QSO survey with a specified band of wave length. However, we should note that the conversion (the bolometric correction) between the magnitude used in observations (e.g., the V-band magnitude $m_v$ or the B-band magnitude $m_b$) and the bolometric magnitude for AGNs is not clear. Thus, it is not worth to argue a detailed estimation of the detection limit on the bolometric magnitude here. 

  At the range with large $z$, say $z>3$, the population of the result plotted as the dot-dash line decreases considerably as $z$ increases, comparing with the previous result (the solid line) in figure \ref{fig:pop1}. This is just as expected, because $L_{lim}$ is an increasing function of $z$ in this case. About the dot-dash line, in the range with small $z$, say $z<2$, the population is over estimated, because $L_{lim}$ which corresponds to a fixed apparent magnitude $m_{lim}$ ($=18.9$ in the case in figure \ref{fig:pop1}) is too small. Such faint nuclei should not to be classified as AGNs. To be more realistic, the detection limit $L_{lim}$ should be switched from (\ref{eq:Llimc}) to (\ref{eq:Llimc2}) at the crossing point ($z=2.5$ in figure \ref{fig:pop1}) of these curves as $z$ increases.

\section{SUMMARY}

The purpose of this work is not to present a realistic model which can precisely  explain the observation, but to demonstrate properties of a fascinating mechanical model, i.e., the Kerr BH fly-wheel model, and to submit a physical scenario of the evolution of QSOs/AGNs not as a speculation, but as a result of the mechanical process. 

We have proposed a magnetohydrodynamic model for `engine' of QSOs/AGNs: the Kerr BH fly-wheel model (see section 3 of this paper and Nitta et al. 1991). This engine is driven by the rotation of BH. The rotation energy of BH is extracted by an electromagnetic process (magnetic breaking). The extracted energy is once stored in the magnetosphere in the form of the Maxwell stress, and then it will produce the plasma outflows. 

One might mislead that the fly-wheel engine is the mechanism only for the radio-loud activity because the released energy produces the outflow. This might cause from a strong impression that highly collimated bipolar jets make the double radio lobes. However, from the observational point of view, the presence of outflows are required for other kind of AGNs. For example, it has been established that, BAL QSOs also have outflows (the ``disk wind'' nearly in the plane of the disk, see Murray et al. 1995). From the theoretical point of view, mechanics of the global structure of outflows is argued in many literature. The produced outflows will show a wide variation of global structures, e.g., the bipolar jets of radio-loud QSOs/AGNs, the equatorial wind of BAL QSOs, or more (see Nitta 1994). Thus we can say that the radio-loud activity is simply one possibility of the fly-wheel engine. 

In the fly-wheel model, we can clarify the properties and the evolution of individual engine (see section 3) parametrized by BH mass $m$, initial Kerr parameter $a$, magnetic field $B_0$ at the source region and a small dimension-less parameter $\epsilon$. These engines are assumed to correspond to QSOs/AGNs. If we obtain the statistical properties for these parameters, we can discuss the statistics of ensemble of QSOs/AGNs. Here we adopt the Press-Schechter formalism for the mass distribution. In the scenario of this work, Kerr BHs are supposed to form at $z=z_{vir}=200$ with nearly maximum angular momentum $a \sim m$ at the formation epoch.  The BH mass is assumed to be 10\% of the total baryonic mass of the proto-galactic cloud. Since the magnetic field $B_0$ seems to be related with the BH mass $m$, we set $B_0 \propto m^\zeta$. As a result, very weak dependence $0 \geq \zeta > -1/2$ is preferred for the consistency with observations. We assume $\zeta=0$ and $B_0=1$ [T] to obtain the figures. The small parameter $\epsilon$ is determined by the physics of plasma injection process, e.g., the pair plasma production or the overflow from the disk halo. Since we do not have widely accepted standard theory of it, we assume as $\epsilon=0.1$ (the distance of the source region is several times the horizon radius). Thus the evolution depends only on the BH mass $m$. 

We have discussed the evolution of the luminosity function and the spatial number density in a period $0 \leq z \leq 5$, and made a qualitative comparison with observations. In the typical case $n=-0.8$ and $\zeta=0$, we obtain the evolution of the luminosity function and the spatial number density with a plausible behavior in the period $0 \leq z \leq 5$ consistent with observations. The brighter-end of the luminosity function is lifted up for $z \geq 3$, then it drops and the curve changes to be short and more steep for $0 \leq z \leq 3$ as shown in figure \ref{fig:LF1}, \ref{fig:LF2}. In accordance with this behavior, the spatial number density evolves as shown in figure \ref{fig:pop1}. 

We should note that these characteristic evolutions obtained in this paper are derived from the evolution of the individual magnetospheric structure in the vicinity of the Kerr BH. This individual evolution is not a speculation but the result based on the MHD picture. In the previous works, e.g., Pei (1995), the evolution of individual AGNs is simply an assumption without any mechanical scenario. We have tried to join the intrinsic Kerr BH magnetospheric physics, i.e., the fly-wheel model with observational facts, and we have succeed to present a mechanical model of the evolution, at least qualitatively. 

In order to explain the observational facts of $0 \leq z \leq 5$, somewhat flat mass function of BHs ($n \sim -0.8$ in equation \ref{eq:IMF}) and a weak dependence of the magnetic field at the source region on the BH mass ($\zeta > -1/2$ in equation \ref{eq:B0}) are preferred. These values of $n$ and $\zeta$ should be determined through an extra physics, however we do not have widely accepted model of them, so we have treated them as free parameters of our picture.

\section{DISCUSSION}

\subsection{Simplifications in our model}

For simplicity, we assumed the BH formation epoch as $z_{vir}=200$ independent of BH mass $m$. From the Compton drag model (see Sasaki \& Umemura 1996), seed BH can formed only at the epoch in which the background photon density is sufficiently high. The formation epoch should be $z > 10^2$. Of course, in the actual case, $z_{vir}$ will depend on $m$. However, when we translate the red shift $z$ to the cosmic time $t$ by virtue of the Einstein-de Sitter model, the variation around $z \sim 200$ corresponds to the order of $\sim 10^6$[yr], and is negligible comparing with the epoch around $z \sim 3$ ($\sim 10^9$[yr]) in which we are interested. Hence, it seems reasonable to suppose that $z_{vir} \sim const.$ independent of $m$. 

We assumed that the dependence of the magnetic field $B_0$ at the source region on the BH mass $m$ as $B_0 \propto m^\zeta$. In the realistic case, $B_0$ should depend not only on $m$ but also on the mass accretion rate. This problem is very difficult and still opened at the present time as discussed in subsection 6.3, however, to discuss this problem as a whole is beyond the scope of this paper. 

We assumed that the initial Kerr parameter $a$ as $a/m \sim 1$. Bi\u{c}\'{a}k \& Dvo\u{r}\'{a}k (1980) showed that extreme Kerr hole does not posses the magnetic field threading the horizon. 
This means that the magnetic breaking process can not extract the rotation energy from extreme Kerr holes. Hence we can not set the initial Kerr parameter as $a/m = 1$. However we should note that exact value of the initial Kerr parameter is not essential. Even if $a/m=$, say, 0.9, 0.5 or 0.3, the explosive epoch $t_{max}$ shifted by only a factor of the order of unity. Such magnitude of ambiguity does not matter in our discussion based on an order estimation. 

The initial mass function or the initial Kerr parameter function of the ``proto-galactic cloud'' have been discussed in some literature (e.g., Sasaki \& Umemura 1996 and Susa et al. 1994), but the general relativistic dynamical process of the BH formation from the proto-galactic cloud has not been solved. The initial Kerr parameter function may not be important comparing with the initial mass function as discussed in above paragraph. Hence we have supposed the initial mass function of the BH from it of the proto-galactic cloud by an assumption that 10\% of baryonic mass collapses to form the seed BH. 

We assumed that the dimension-less small parameter $\epsilon \equiv m \Omega_F=0.1$ in the calculation. This value corresponds to the situation that the source region (the plasma injection region or the pair creation region) is located at a radius several times the horizon radius ($\sim 4.6 m$ for $\epsilon=0.1$). This might be plausible for the outer gap model of the pair creation. The factor $\Omega_F$ of the definition of $\epsilon$ roughly coincides with the Keplerian angular velocity at the source region. If the locus of the plasma source is fixed at a radius several times the horizon radius from the outer gap model, $\Omega_F$ depends only on $m$. In the evolutionary model of Nitta et al. (1991), $m \sim const.$ during the characteristic time scale of the angular momentum extraction which is defined as the time scale of the evolution. Hence we can assume that $\Omega_F \sim const.$, thus $\epsilon \sim const.$ independent of the time. 

In our model, each BH suddenly ceases to release energy at $t=t_{max}$ corresponding to the situation $\omega_H=\Omega_F$. If $\Omega_F$ of a BH magnetosphere distributes in a range, say, $\Omega_{1} \leq \Omega_F \leq \Omega_{2}$ as a function of the magnetic flux function, the fly-wheel activity gradually ceases on a field line in order of decreasing value of $\Omega_F$. If this variety of $\Omega_F$ does not result the vary $\epsilon$ in order of magnitude, it does not affect our qualitative discussion. 

In order to relate the time after the formation of blackholes and the redshift, we need a cosmological model. In the main discussion of this paper, we adopt the Einstein-de Sitter model (see eq. \ref{eq:t-z}) for simplicity. Of course, the Einstein-de Sitter model is very classical, and recent observational studies of cosmology support the model with the cosmological constant, i.e., the Lemaitre model. Here we have to estimate the difference of results between the cases with the Einstein-de Sitter model and the Lemaitre model. We provide the look-back times $t_1(z)$ and $t_2(z)$ of the epoch with the redshift $z$ in the Einstein-de Sitter model and the Lemaitre model, respectively. If we choose the parameters $\Omega_0=0.1$ and $\lambda_0=0.9$ in the Lemaitre model, the look-back time of the blackhole formation epoch ($z=10^2$) can be estimated as $t_2=1.92 t_1$. At the epoch $z=5$ in which we are interested here, $t_2=1.83 t_1$. These differences are simply in the factor of the order of unity. Hence the difference of the results between these two cosmological models does not matter in our discussion because the evolutionary process of our discussion (see section 3) is based on an order estimation.

\subsection{2-types of BH engines}

In this paper, our attention is focused to the fly-wheel model. However, BH-accretion disk systems also seem to include another type of engine: the fuel (accretion powered) engine. The elementary process of the fuel engine is release of the gravitational energy of the accreting matter, so that the activity strongly depends on the mass accretion rate. It is widely believed that the accretion rate is roughly determined by the Eddington limit. This idea is based on a speculation that the regulating stage of the entire accretion process is the final stage, i.e., the accretion onto the BH. However, we do not have any theoretical conviction that how the entire system (i.e., a galaxy) determines the accretion rate. In another word, how the system can remove the angular momentum of the accreting matter to realize such accretion rate. 

In order to determine the activity of the fuel engine, we must solve the extraction of the angular momentum in a very wide spatial range. If we want to obtain the accretion rate of a range from an angular momentum extraction mechanism, we need the accretion rate outside this region where another mechanism may regulate the accretion rate, and so on. This endless chain seems to be hopeless to solve completely. Thus, it is so difficult to assemble the angular momentum extraction mechanisms of each range into a consistent theory, that we do not have another way except to treat the final accretion rate onto the BH as a free parameter. This is the point of difficulty to make evolutionary scenario of QSOs/AGNs based on the fuel model. 

In the model adopted in this paper, we do not consider the activity of the fuel engine at all. However, the author believes that coexistence of these 2 types of engines is undoubted. Even after the fly-wheel engine ceases to release energy, the fuel engine continues to act while the accretion is continued. If we make an appropriate assumption to estimate the luminosity due to the fuel engine, we should add the contribution of the fuel engine. For simplicity, if we assume a constant energy release of the fuel engine, i.e., the Eddington luminosity (depends only on $m$), this luminosity is comparable with the initial luminosity of the fly-wheel engine. In this model, the fuel engine works constantly and the fly-wheel engine works in a period $t_{max}$ after the BH formation. Even after the fly-wheel engine dies, the luminosity does not vanish, but decreases to the Eddington luminosity. 

In section 2, the fly-wheel model is characterized as that the mechanism of it can be clarified by the closed discussion in the vicinity of the BH. However this statement might be somewhat exaggerate. The fly-wheel engine seems to relate with the fuel model at the point that the magnetic field $B_0$ at the source region may depend on the accretion rate. $B_0$ is provided as the magnetic field strength averaged in a larger macroscopic scale than the scale of turbulence generated in the accretion disk. Such large scale magnetic field should be amplified by the accretion. The accretion plasma carries the frozen magnetic field into the inner magnetosphere and compresses it. Against this process, small but finite resistivity dissipates the magnetic field. Then the saturated level of the magnetic field strength is determined by the equilibrium of the compression and the dissipation. Unfortunately, this problem has not been solved yet as in the next subsection. Hence we have assumed that $B_0$ depends on the BH mass $m$ like equation (\ref{eq:B0}) because the accretion rate seems to depend on $m$.

\subsection{Ambiguity of the fly-wheel power estimation}

Ambiguities are still remained on the estimation of the power of the fly-wheel activity mainly due to the following two reasons. The first is due to some theoretical ambiguities of the estimation of the magnetic field strength near the BH. The second is ambiguity of the innermost magnetospheric structure of the BH-accretion disk systems, especially whether the field lines threading the horizon (or the innermost region of the accretion disk) are open toward the infinity or closed to be loop-like one. 

The poloidal magnetic field strength is traditionally estimated by an intuition of a principle of equipartition between the magnetic energy density and the gravitational one or the thermal one (see, e.g., Shakra \& Sunyaev 1973). Recent numerical studies based on nonlinear evolutionary process of the resistive MHD seem to support the result from the equipartition. For example, Matsumoto et al. (1997) conclude that predominantly toroidal magnetic field is amplified by a differential rotation of the disk and the plasma $\beta$-value of $\beta \sim 10$ can be achieved. If there is significant magnitude of the poloidal magnetic field, the saturation level will be more large and $\beta \sim 1$ might be achieved. 

While this conjecture of equipartition is now widely accepted, there is still room for a disagreement about this point. Recently Livio et al. (1999) critically assess the efficiency of the Blandford-Znajek (BZ) process (the magnetically dominated case of the BH fly-wheel model) comparing with other disk activities. They reconsider the field strength in the innermost region of BH-disk system. In their result, power of the BZ process is dominated by the fly-wheel or the fuel (viscous heating) power of the innermost region of the accretion disk. The problem of the saturated strength of the poloidal magnetic field is still an open question. 

However we should note that the energetics of the fly-wheel process depend not only on the strength of the magnetic field but also on the inner magnetospheric structure of the BH-disk systems. The extracted Poynting energy flux due to the fly-wheel process is carried along the poloidal magnetic field lines, and will be converted to the kinetic energy of plasma outflow at some distant region from the horizon. Hence only open magnetic field lines can take place to the energy extraction toward very distant region. For example, Nitta et al. (1991) give schematic figure for the innermost magnetospheric structure (see figure 3 of that paper). In their result, the magnetic field lines connecting to the innermost region of the disk are closed (a loop-like structure connecting the BH and the disk), and do not contribute to the energy extraction. Open field lines are emanated from a high latitude region of the BH and the outer part of the disk. In this case, the discussion of Livio et al. (1999) should be altered.  

Thus efficiency of the fly-wheel process is closely combined with the disk dynamo process and the magnetospheric structure of the innermost region. These are very important but still open questions in the current state, and to argue this point would carry us too far away from the purpose of this paper.

\subsection{Discrimination of QSOs}

As discussed in section 4, in the fly-wheel model, we can only estimate the total output power of the engine, and we cannot discuss the spectrum of resultant radiation. Hence the only way to distinguish QSOs/AGNs from normal galaxies is setting a criterion, say $L_{lim}$, on the bolometric luminosity. Here we suppose that entire released energy is perfectly converted to radiation, and suppose that the engines having the luminosity greater than $L_{lim}$ can be treated as QSOs/AGNs. This simplified procedure is obviously far from the actual QSO number counting studies. In future study, the problem of the resultant spectrum of the radiation should be solved. This is possible only if we solve the physics of plasma outflows being generated by the fly-wheel engine. This is, needless to say, one of the most difficult open questions in the magnetospheric astrophysics. 

From figure \ref{fig:pop1}, the locus of the peak of population strongly depends on the criterion $L_{lim}$. If we set the smaller $L_{lim}$, the peak shifts to the smaller $z$. This means that if we survey AGNs in more deep, we will find more and more faint AGNs including low mass BH. These may correspond to Seyferts. However we should note if the mass of central BH is too small, the nucleus activity is dominated by its host galaxy. Such objects may not be classified as QSOs/AGNs. In this meaning, the criterion $L_{lim}=10^{38.9} \ \mbox{or}\ 10^{37.7}$[W] adopted in this paper might be plausible, because these values dominate the typical luminosity of normal galaxies $10^{37}$[W] (Andromeda galaxy).

\subsection{Similarity of radio-loud and radio-quiet AGNs}
Similarity among all kinds of AGNs is widely accepted from the observational point of view. The spectral energy distribution (SED) of radio-loud AGNs and radio-quiet AGNs are quit similar except the radio range (see, e.g., Elvis et al. 1994). We also cannot find any intrinsic difference in the evolution of the spatial number density of the optically selected QSOs, the flat-and-steep spectrum sources and radio-loud QSOs (see Shaver et al. 1996).  These observational evidences of similarity implicitly suggest the universality of physical process of QSOs/AGNs. 

As mentioned in section 5, the author believes that the fly-wheel model is applicable not only for radio-loud AGNs. The variation of AGNs might be caused from the variation of proper parameters (BH mass, BH angular velocity, magnetic field strength, etc.). The difference of parameters will lead to different structure of outflows (see Nitta 1994). We may expect that outflows having different structure will produce different types of spectrum of the radiation. However, the correspondence of a global structure of the outflow and a resultant spectrum of radiation is still unclear. This should be a future problem.

\subsection{Other stories of the fly-wheel activity}

There are other stories to make Kerr BH as central BH of QSOs/AGNs. Here we mention two of them: the merging of BHs and the spins up by accretion. 

Wilson \& Colbert (1995) discussed the formation of AGN BHs by merging process. They tried to explain the difference between radio-loud and radio-quiet AGNs. As well known, number fraction of radio-loud AGNs is only 10\% of total AGNs, and radio-loud galaxies are mainly observed as elliptical galaxies. From these points, they supposed that radio-loud AGNs are merger events of BHs. Merging of 2 galactic nuclei produce a quickly rotating Kerr BH, and a period after merging, the Blandford-Znajek process acts and shows radio-loud activity. Similarly, Moderski \& Sikora (1996) supposed to make quickly rotating BH by very large mass accretion. 

The scenarios discussed in these papers are alternative one. There is room for another possibility that the fraction of radio-loud to radio-quiet may be related to probability to make jet-like outflow. The fly-wheel engine can form various structures of outflows (see Nitta 1994). If well-collimated jet-like structure which is nearly perpendicular to the galactic disk is formed, this will be observed as radio-loud one (or the radio galaxy, see Urry \& Padovani 1995), because the terminal shock in the jet will locates far from the galactic disk due to very low ambient matter density in this direction. If equatorial wind is formed, this will be BAL QSO (see, e.g., Cohen et al. 1995 or Murray et al. 1995 ). Of course, these are simply speculations, because we do not have widely accepted physics for the structure formation of plasma outflows yet. Anyway, since we do not have authorized theory, we must try to test various possibilities. 

In some literature, the fuel process (accretion from disk) and the fly-wheel process (Blandford-Znajek process) are simultaneously considered (see, for example, Moderski \& Sikora 1996 and Ghosh \& Abramowicz 1997). In these models, the accretion contributes to spin up the BH, but the BZ process suppresses it. On the contrary, Nitta et al. (1991) imply that MHD accretion onto the Kerr BH extracts the angular momentum and spins down the rotation of the Kerr BH when $\Omega_F < \omega_H$ (see section 3). This is a natural result of MHD accretion onto the Kerr BH (see Takahashi et al. 1990).  

One might think this is contrary each other. However the author does not think so. Moderski \& Sikora (1996) and Ghosh \& Abramowicz (1997) suppose to start with slowly rotating BH ($\Omega_F > \omega_H$), on the contrary, Nitta et al. (1991) suppose quickly rotating BH ($\Omega_F \ll \omega_H$). This difference causes from the difference of concepts of the models. However, we should note that this difference is essential. 

In our model, the energy source of the Kerr BH fly-wheel engine is inherently obtained rotation energy of the central BH. The origin of this energy is the tidal interaction during the collapse of the proto-galactic cloud with other density fluctuations. Once the fly-wheel engine starts to act, the rotation energy monotonically decreases, and stops at the state $\Omega_F=\omega_H$. 

On the other hand, in other models, accretion energy converts to the fly-wheel type activity, and they relate to the radio-loud activity.  For example, central BH is normally in a state of slow rotation, however, if coalescence of BHs (Wilson \& Colbert 1995) or very large mass accretion (Moderski \& Sikora 1996) occurs, central BH spins up and the fly-wheel engine starts to wok. In these models, the energy source is, consequently, the accretion energy.

\subsection{Dormant quasar: Fornax A}

Let us notice a splendid example Fornax A (NGC1316, see Iyomoto et al. 1998) which seems clearly show properties of the fly-wheel engine. In general, we cannot observe evolution of a galaxy because of very long lifetime of it, however Fornax A is a particular case which we can obtain the evidence of the evolution for recent 0.1 Gyr. This is a radio galaxy with double radio lobes. The nucleus should be active ($> 4 \times 10^{34}$[W] in 2-10 keV X-ray luminosity) at least 0.1 Gyr ago, while the present activity is `dormant' ($2 \times 10^{33}$[W] in 2-10 keV X-ray luminosity). 

We can guess the reason of it based on the fly-wheel model as follows. The fly-wheel engine was still active 0.1 Gyr ago, and the nucleus ejected plasma outflows (bipolar jets) and made the radio lobes. At an epoch within a past 0.1 Gyr, the fly-wheel engine ceased to work and nucleus became to be dormant. The radio lobe can emit radiation within a period determined by the Synchrotron cooling time without the energy supply by the fly-wheel engine. In this sense, the fly-wheel engine of Fornax A is not `dormant' one but `dead' one if without some mechanisms to spin up the central BH again. However the fuel engine can act after the fly-wheel activity cease. This corresponds to the present nucleus activity. In this case, the peak fly-wheel activity is an order of magnitude greater than the fuel activity. This is just a result of our model with $\epsilon \sim 0.1$. 

\vspace{0.5cm}

{\bf Acknowledgement}\\

The author wishes to thank Drs. K. Aoki, K. Okoshi, S. Satoh, S. Kameno, T. Yamamoto and T. Totani, and Mr. Y. Tutui at National Astronomical Observatory of Japan for helpful comments and criticisms. The author also thanks to the anonymous referee for comprehensive comments for improvement. The author also thanks to Mr. S. Abe and Dr. A. Kawamura for the technical support.

\end{document}